\documentclass[%
aip,
amsmath,amssymb,
reprint,%
]{revtex4-1}

\usepackage{graphicx}
\usepackage{dcolumn}
\usepackage{bm}
\usepackage{braket}
\usepackage[utf8]{inputenc}
\usepackage[T1]{fontenc}
\usepackage{mathptmx}
\usepackage{siunitx}
\usepackage{url}
\usepackage{subcaption}
\usepackage{hyperref}
\usepackage[dvipsnames]{xcolor}

\def\Put(#1,#2)#3{\leavevmode\makebox(0,0){\put(#1,#2){#3}}}

\begin{document}
	
	
	\title[6 state-4 state protocol]{Demonstration of a 6 State-4 State Reference Frame Independent Channel for Quantum Key Distribution}
	
	\author{Ramy Tannous}
	\email{ramy.tannous@uwaterloo.ca.}
		\affiliation{ 
		Institute for Quantum Computing, Department of Physics and Astronomy, University of Waterloo, Waterloo, Ontario, N2L 3G1 Canada
	}
	
	\author{Zhangdong Ye}
		\affiliation{ 
		Institute for Quantum Computing, Department of Physics and Astronomy, University of Waterloo, Waterloo, Ontario, N2L 3G1 Canada
	}
	\affiliation{ 
		State Key Laboratory of Low-Dimensional Quantum Physics and Department of Physics, Tsinghua University, Beijing 100084, China 
	}
	\author{Jeongwan Jin}
		\affiliation{ 
		Institute for Quantum Computing, Department of Physics and Astronomy, University of Waterloo, Waterloo, Ontario, N2L 3G1 Canada
	}
	\author{Katanya B. Kuntz}
		\affiliation{ 
		Institute for Quantum Computing, Department of Physics and Astronomy, University of Waterloo, Waterloo, Ontario, N2L 3G1 Canada
	}
	\author{Norbert L{\"u}tkenhaus}
		\affiliation{ 
		Institute for Quantum Computing, Department of Physics and Astronomy, University of Waterloo, Waterloo, Ontario, N2L 3G1 Canada
	}
	\author{Thomas Jennewein}
		\affiliation{ 
		Institute for Quantum Computing, Department of Physics and Astronomy, University of Waterloo, Waterloo, Ontario, N2L 3G1 Canada
	}
	\affiliation{ 
	Quantum Information Science Program, Canadian Institute for Advanced Research, Toronto, Ontario, M5G 1Z8 Canada
}

	\date{19 November 2019}
	
	\begin{abstract}
		We study a reference frame independent (RFI) quantum key distribution (QKD) protocol using six states for Alice and only four states for Bob, while previous RFI protocols require a six state analyzer for Bob. Our protocol can generate a secure key for any possible phase of the entangled state, provided the variation is small compared to the measurement rate as shown by our numerical key rate analysis. We perform a proof-of-principle experiment using polarization entangled photon pairs. In the presence of a varying rotational phase, we obtain a consistently low error rate of less than $4\%$, indicating the feasibility of this protocol for QKD. Our RFI protocol is hence beneficial but not limited to applications in satellite or mobile free-space QKD, where a communication node must limit resources and restrict the number of measured states to four instead of six.
	\end{abstract}
	
	\maketitle

	Quantum key distribution (QKD) protocols provide a means of generating and sharing an encryption key between two parties, Alice and Bob, with the security guaranteed by the laws of quantum physics\cite{QKD_Security}. There is an on-going effort to improve the practicality of QKD \cite{QKD_Security}. In many protocols, Alice and Bob need continuous agreement of all shared measurement frames during the entire communication period\cite{BB84,BBM92,6-state}. The definition of the measurement frames is essential, particularly for protocols that use polarization encoding where a geometric reference is required. However, this demand can be relaxed by employing reference frame independent (RFI) protocols, which allow for all or some of the measurement frames to rotate by a slowly varying relative phase\cite{laing2010reference,SPEDALIERI2006340,beheshti2019entanglement,thinh2012tomographic}. We present and implement a RFI 6 state-4 state protocol that uses polarization entangled photons, where one receiver (Bob) can only perform measurements in two Pauli bases, while the other receiver (Alice) can measure in the usual three Pauli bases. Despite the reduced measurement at Bob, we demonstrate that the protocol is still RFI and suitable for QKD.
	
	We apply our protocol to the scenario where a quantum source is connected to a free space quantum channel by an optical fiber that is attached to a transmitter telescope. Typically the optical fiber will cause equal rotations to all polarization bases due to manufacturing tolerances, and thermal and mechanical stress. We use polarization maintaining fibers (PMF) since they were developed to specifically combat these rotations\cite{noda1986polarization}. However, any polarization not aligned to one of the two principle axes (slow or fast axis) will be subject to a large birefringence, which typically causes entangled photons to decohere. Therefore, PMF emulates the slow drift for two bases, while the third basis is fixed and therefore provides a good experimental platform for testing the protocol. 
		
	
	For the 6 state-4 state RFI QKD protocol, Alice and Bob share an entangled state $\rho_{AB}$, each making Pauli measurements on half of the two-qubit state. Although here we use entanglement, we note that this protocol can also be used in other QKD implementations, including prepare and measure schemes. In our experiment, the computational basis is the horizontal-vertical polarization basis ($\sigma_z$), which is the fixed basis. Alice applies a six state measurement on her qubit, measuring in the $\sigma_z\text{, } \sigma_x\text{ and }\sigma_y$ bases, while Bob applies a four state measurement on his qubit in the $\sigma_z\text{ and } \sigma_x$ bases. The reduced measurement at one receiver brings several benefits. Most importantly, it reduces the resources required for the state analyzer, which is vital for receivers that have resource constraints, such as satellites like the Quantum Encryption and Science Satellite\cite{QEYSsat}.
	
	The omission of the third basis requires adjusting the parameters used in the symmetric 6 state protocol presented in Laing et al.\cite{laing2010reference}. We therefore define a phase independent parameter ($C$)
	\begin{equation}
	\label{eq:myc}
	C=\sqrt{\braket{\sigma_x\otimes \sigma_x}^2+\braket{\sigma_y\otimes \sigma_x}^2},
	\end{equation}
	where $\braket{M}$ is the expectation value of the two qubit positive-operator valued measure (POVM) $M$, defined as,
	\begin{equation}
	\label{eq:exp}
	\braket{M}=\frac{M_{++}-M_{+-}-M_{-+}+M_{--}}{\sum_{ij}M_{ij}}.
	\end{equation}
	$M_{ij}$ $(i,j=+,-)$ are the coincidence counts of the results for the POVM $M$, and $\sum_{ij}M_{ij}$ is the total coincidence counts measured of $M$. By Pauli algebra, we see that $C\leq 1$, with the equality corresponding to a maximally entangled state. $C$ provides a second basis that can be useful in a quantum communication context, which we will call the "diagonal*" basis. $C$ is used to monitor the quality of the quantum channel and any significant drop from unity can be attributed to an eavesdropper's intervention\cite{laing2010reference}. In addition, $C$ is constant in the presence of an arbitrary relative phase $\phi$ between the computational states, provided $\phi$ remains constant over each individual finite measurement interval. For example, with the idealized Bell state that is subject to $\phi$,
	\begin{equation}
	\label{eq:bell}
	\ket{\Psi}=\frac{1}{\sqrt{2}}(\ket{H}_A\ket{V}_B+ e^{i\phi}\ket{V}_A\ket{H}_B),
	\end{equation}
	we find $\braket{\sigma_x\otimes \sigma_x}=\cos\left(\phi\right)$ and $\braket{\sigma_y\otimes \sigma_x}=\sin\left(\phi\right)$. Eq.~\ref{eq:myc} then obtains $C=(\cos\left(\phi\right)^2+\sin\left(\phi\right)^2)^{1/2}=1$, regardless of $\phi$, thus confirming the integrity of the quantum channel. For additional comments on the security see Laing et al.~\cite{laing2010reference}.
	
	The quantum bit error ratios (QBER) in the computational basis and the "diagonal*" basis are given by,

	\begin{eqnarray}
	\label{eq:QBer}
	\text{QBER}_{HV}=\frac{1-\braket{\sigma_z\otimes \sigma_z}}{2}, \hspace{2mm} \text{QBER}^*_{Diag}=\frac{1-C}{2}.
	\end{eqnarray}	
	The $\text{QBER}^*_{Diag}$ is an effective QBER which monitors any drop in $C$. For a more in-depth analysis of the QBER for RFI protocols see Yoon et al.\cite{yoon2019experimental}. We estimate an asymptotic key rate from the estimated QBER by\cite{QKD_Security},
	\begin{equation}
	\label{eq:keyr}
	K \geq qR[(1-f H_2(\text{QBER}_{HV})-H_2(\text{QBER}^*_{Diag}))]
	\end{equation}
	where $q$ is the basis reconciliation factor, ($1/6$ in our case), $R$ is the coincidence detection rate, $H_2(x)$ is the binary entropy function, and $f$ is the bidirection error correction efficiency\cite{brassard1994lecture,lutkenhaus2000security}. We assume error correction at the Shannon limit so $f = 1$. The analytical key rate of Eq.~\ref{eq:keyr} does not account for individual detection efficiencies, nor any multiphoton contributions. Note, this 6 state-4 state protocol will yield the same key rate as the 6 state-6 state protocol for unitary channels, i.e. rotations. Later we take an in depth look at the key rate estimation using numerical methods.
	
\begin{figure*}[htpb]
		\raggedright 
		\includegraphics[width=15cm]{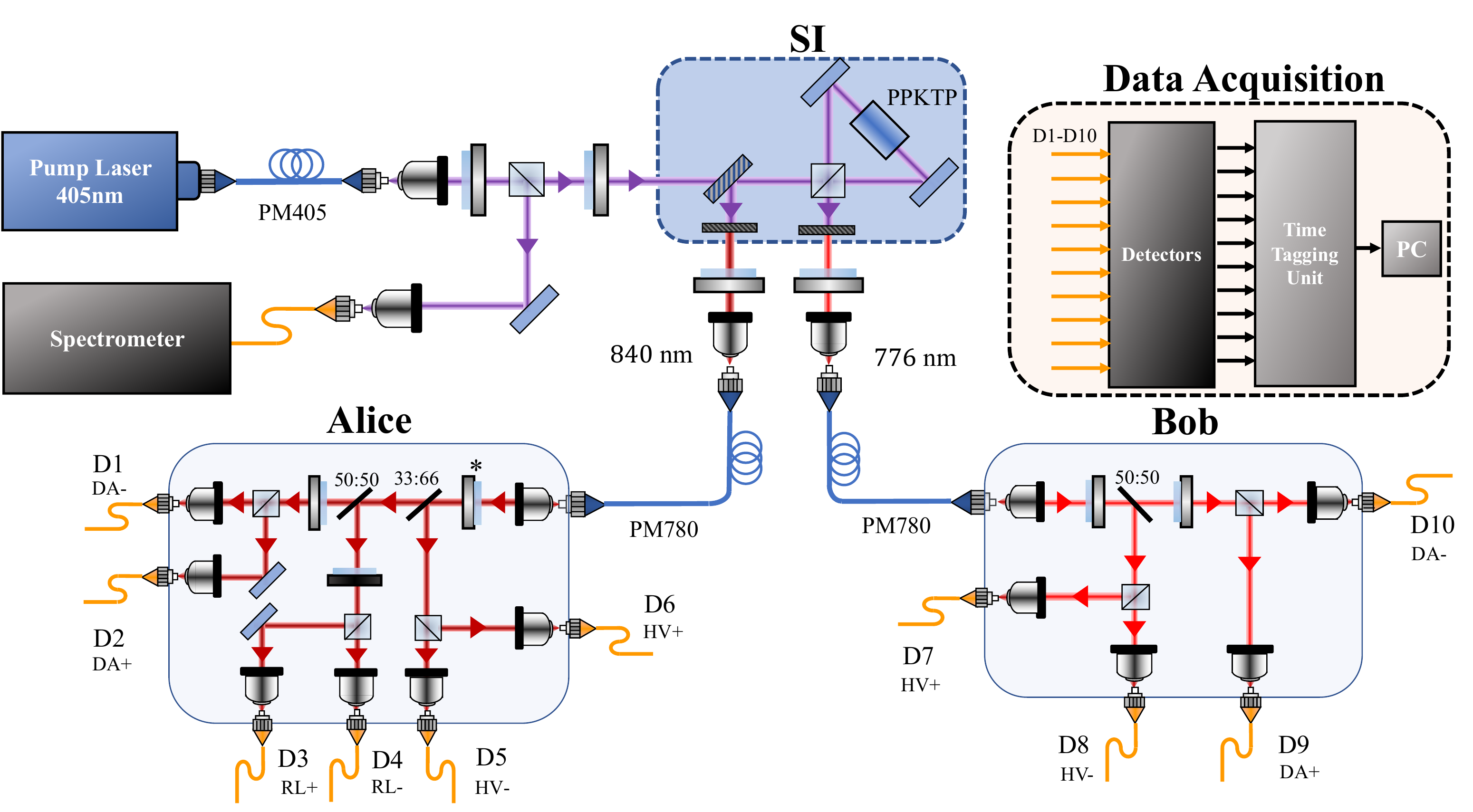}\llap{\Put(50.5,222.1){\mbox{\includegraphics[trim={7cm 0cm 22.4cm 0cm},clip,height=4.5cm]{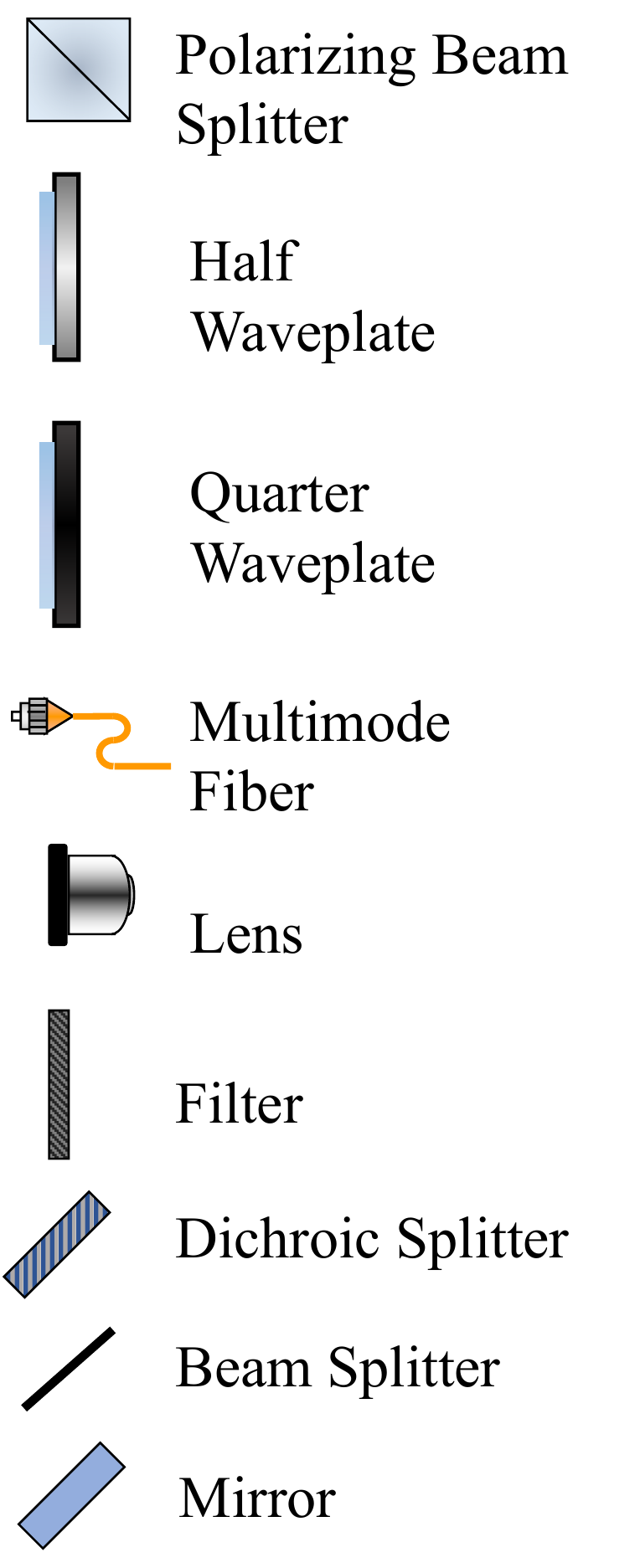}}}}
		\caption[RFI Sagnac]{Experimental setup. A \SI{405}{nm} laser pumps a type-II periodically poled potassium-titanyl phosphate (PPKTP) non-linear crystal in a Sagnac interferometer\cite{Sagnac} (SI). Entangled \SI{776}{nm} and \SI{840}{nm} photons are collected into polarization maintaining fibers (PM780-HP). Alice performs a six state measurement on the \SI{840}{nm} photon, while Bob performs a tomographically incomplete four state measurement on the \SI{776}{nm} photon. Ten silicon avalanche photodiodes are used to detect the photons. Coincidence and single events are recorded and analyzed by a time-tagging unit and a computer. A spectrometer is used to monitor the pump spectra during the measurements. * indicates the half-wave plate rotated about its vertical axis used to manipulate and control the external the phase variations.}
		\label{fig:app_rfiSagnac}
	\end{figure*}

	
	The entangled photons used in the experiment are created using a Sagnac interferometer\cite{Sagnac} that produces type-II spontaneous parametric down conversion with the signal photons at \SI{776}{nm} and the idler photons at \SI{840}{nm}, see Fig.~\ref{fig:app_rfiSagnac}. The down converted photons are then collected into PMF with the horizontal and vertical polarizations aligned to the two principle axes of the PMF, while any other polarization not aligned to these axes will experience an arbitrary rotational phase. The difference in group velocity between the slow and fast components of the polarization causes temporal displacement, or "walk-off", which we force to be symmetric by sending the entangled photons through \SI{2}{m} of 780~\si{nm} PMF. The symmetry ensuring that the walk-off induced to each photon is approximately equal. In addition, this walk-off must be less than the pump's coherence time to avoid any distinguishability of the photon pairs, since the coherence times of the entangled photons are transferred from the pump\cite{shtaif2011nonlocal,Transfer_Kulkarni:17}. For example, in our experiment the fibers induce a walk-off of approximately \SI{2.34}{ps} which is much less than coherence time of the pump, \SI{1.08}{ns}.
	
	The signal photons are measured by Bob using a four state polarization analyzer, while the idler photons are measured by Alice using a six state polarization analyzer. All single photon and coincidence counts are measured and recorded using silicon avalanche photodiodes (Excelitas) that have a timing resolution of \SI{500}{ps}, and a time tagging unit (Universal Quantum Devices). The coincidences are measured using a \SI{1}{ns} correlation window, and accumulated over a \SI{1}{s} integration time. It is important to select a correlation window that is larger than the detector resolution as the temporal uncertainty of a photon's arrival is limited by the detector timing jitter. In addition, care is taken to ensure that the basis selection in each state analyzer is unbiased by balancing the optical path efficiencies within each state analyzer. However, the detection efficiencies may vary between the different detectors, and knowledge of these discrepancies is a critical component of the security analysis to follow. Overall, the resulting entangled qubit state (ignoring the vacuum component) can be approximated to Eq.~\ref{eq:bell} with $\phi$ being the phase accumulated from the relative phase between the slow and fast axes of the PMF, Alice and Bob's optical elements, and the phase of the pump laser.

\begin{figure*}[htpb]
	\centering
	\begin{subfigure}[t]{0.45\textwidth}
		\centering
		\includegraphics[width=\textwidth]{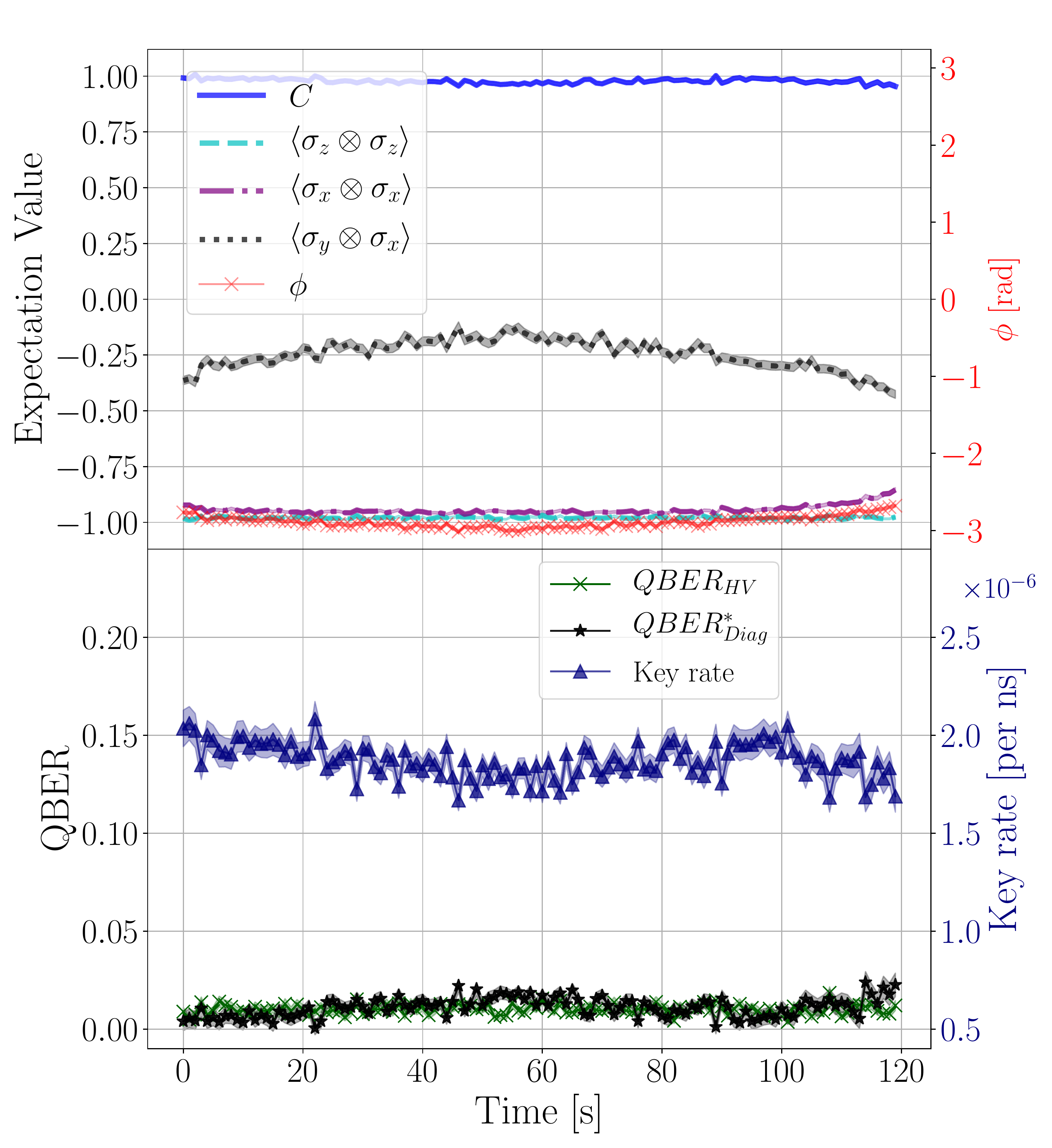}\llap{
			\parbox[b]{15.7cm}{\large{(a)}\\\rule{0ex}{8cm}
		}}
		
	\end{subfigure}%
	\hspace{1mm}
	~ 
	\begin{subfigure}[t]{0.45\textwidth}
		\centering
		\includegraphics[width=\textwidth]{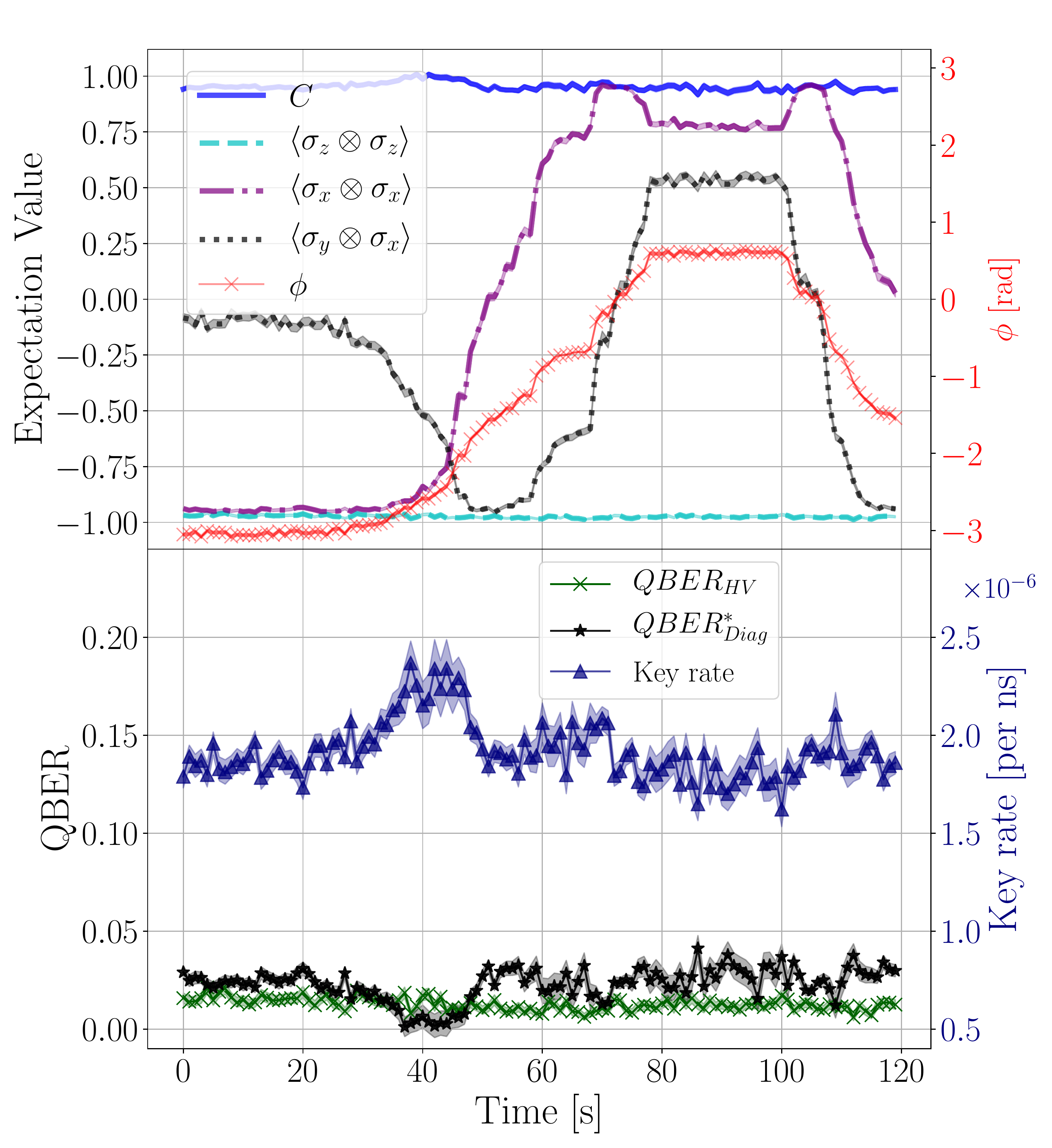}\llap{
			\parbox[b]{15.7cm}{\large{(b)}\\\rule{0ex}{8cm}
		}}
	\end{subfigure}
	\caption[Caption for LOF]{Upper: the experimental expectation values and phase $\phi$. Bottom: QBER and key rate during the phase change. (a) System left undisturbed. The average $C$ is $0.97(1)$. The average $QBER^*_{Diag}$ is $0.0112(4)$ and total QBER is $0.021(6)$. The average estimated key rate is $1.87(8)\times10^{-6}$. (b) Varying phase induced by a HWP in Alice's analyzer Fig.~\ref{fig:app_rfiSagnac}. The average $C$ is $0.96(4)$. The average $QBER^*_{Diag}$ is $0.022(2)$ and total QBER is $0.03(1)$. The average estimated key rate is $1.9(1)\times10^{-6}$. The shaded regions represent the calculated error of the respective value. Error bounds are present in all figures, however, some are too small to be visible. The error bounds are derived by propagating statistical counting errors. The expectation values, QBER and normalized key rates are calculated from Eq.~\ref{eq:exp}, Eq.~\ref{eq:QBer} and Eq.~\ref{eq:keyr} respectively.}
	\label{fig:res}
\end{figure*}

	We report the results of two experimental conditions. In the first scenario (a), the system is undisturbed such that the rotational phase is being caused by the birefringence of the PMF, and any phase change is attributed to the physical or thermal stress of the fiber. In the second scenario (b), the phase changes are introduced using an additional birefringent element\footnote{In (b) a half-wave plate (HWP) rotated about its vertical axis with it's fast axis aligned to the horizontally polarized photons is used to induce a phase. It is the first HWP found after the output on Alice's side, see * in Fig.~\ref{fig:app_rfiSagnac}.}. From the collected coincidence measurements, we compute the expectation values of all the possible POVMs using Eq.~\ref{eq:exp} and calculate $C$. Indeed for both scenarios, $C$ appears to be a constant function of the phase, as shown in Fig.~\ref{fig:res}. However, in Fig.~\ref{fig:res} (b) we see some drop of $C$ which can be attributed to additional photon adsorption due to the additional birefringent element.


Using the correlation data, we estimate the QBER according to Eq.~\ref{eq:QBer}. Even with the presence of a random relative phase, a low overall QBER is maintained with an average total QBER of $0.021(6)$ for (a) and $0.036(8)$ for (b). A total QBER of less than 0.11 is typically required to perform a secure key transfer for systems based on qubits. Our observed QBER are well below this threshold indicating that this protocol is robust to phase drifts despite the lack of tomographically complete measurements. Given the low overall QBER, asymptotic normalized key rates are estimated for both scenarios using Eq.~\ref{eq:keyr}, the results are shown in Fig.~\ref{fig:res}. The average key rate per time slot (1ns) for (a) is $1.87(8)\times10^{-6}$ and $1.9(1)\times10^{-6}$ for (b)\footnote{The spike in the key rate around 40s in (b) is attributed to imperfections in the projective measurements, and variations in the coupling efficiencies caused by changing the phase of the state. Both caused a change in the expectation values that increased $C$. Furthermore, the spike indicates how a slight change of detection efficiencies could impact the key rate, which future security analyses will need to take in consideration as unreliable measurement devices.
}, while the theoretical limit of our system given by Eq.~\ref{eq:keyr} is $0.167$. The large discrepancy from the theoretical limit is due to the enormous amount of clock slots, which accordingly shrinks the key rate. Effectively, we assume that the experimental set-up runs on a clock given by the time scale of the coincidence detection electronics (\SI{1}{ns}).

Beyond the analytically calculated key rate of Eq.~\ref{eq:keyr}, we examine the key rate by taking into account the effects of all the different detection efficiencies. We implement a detailed modeling of the physical set-up, and perform a numerical security analysis along the lines of Winick et al. \cite{Winick2018reliablenumerical}. 

To accomplish this, we follow three steps. In the first step, we analyze the data to find self-consistent values of the detection efficiencies for the various polarization detection paths. We fit the experimental data with a quantum optical model of the source and detection that is restricted to vacuum and single photon signals. We verify the self-consistency of this model, use the combined data from all experimental runs, and find fit parameters for the detection efficiencies (ranging between $0.063-0.120$). In a more rigorous security analysis, one must determine the detection efficiencies via independent calibration measurements that cannot be externally influenced or correlated. However, a rigorous finite size security analysis is beyond the scope of this work.

In the second step, we deal with the fact that the experiments provide frequencies instead of probabilities of observed events since our asymptotic key rate calculation requires probabilities of events. We therefore use a maximum likelihood method (MLM)\cite{teo2013informationally,RrhoR,Superfast}$^,$\footnote{Our selection of the MLM is not restrictive and we could chose other tomography methods} to reconstruct a joint, two-party density matrix in the respective vacuum and single-photon subspaces. We use a model of the POVM elements, including the detection efficiency parameters from step 1, to extract probabilities for the observed events. To facilitate step 2, we impose a time interval structure on the data to capture the effect of vacuum detections that are predominant due to the enormous amount of clock slots from the effective experimental clock of \SI{1}{ns}. 

\begin{figure}[htpb]
	\centering
	\includegraphics[width=0.46\textwidth]{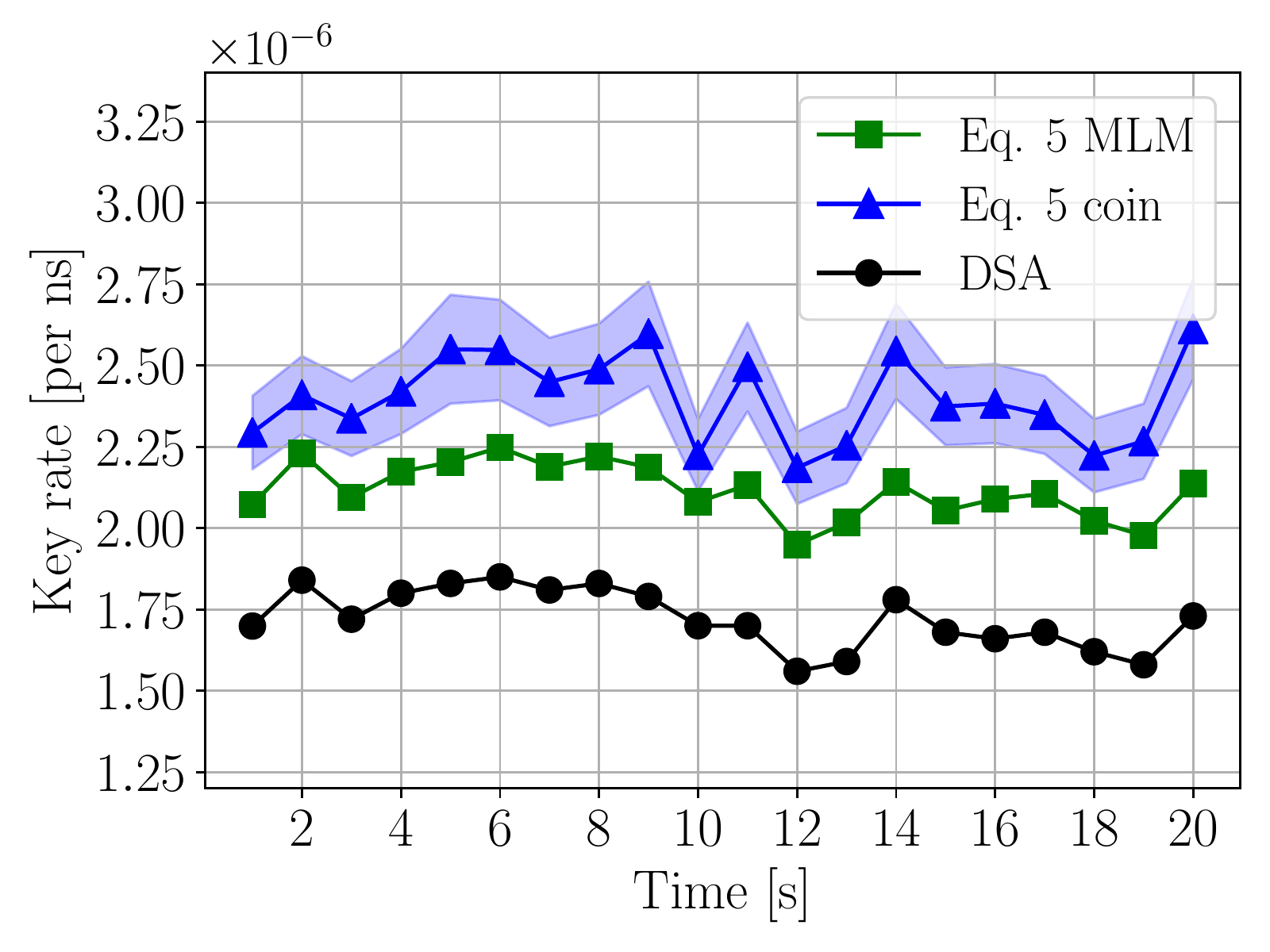}
	\caption{Normalized key rates for 20 seconds of experimental data. Black circles: the three step detailed security analysis (DSA). Green squares: Eq.~\ref{eq:keyr} where the QBER is computed using the MLM estimated density matrices. Blue triangles: Eq.~\ref{eq:keyr} where the QBER is computed using the coincidence counts and Eq.~\ref{eq:QBer}. The error bounds are derived from statistical counting errors. No error analysis is provided for the values derived from the MLM. The discrepancies between the squares and the triangles can be attributed to the MLM used. However, discrepancies between the circles and the squares is independent of the state estimation method.}
	\label{fig-20keyrate2}
\end{figure}

In the third step, using the determined detection efficiencies and derived observation probabilities, we perform an asymptotic numerical security analysis, following the procedures that are described in Winick et al.\cite{Winick2018reliablenumerical}. We show the resulting key rates in Fig.~\ref{fig-20keyrate2}, where we can see the range of asymptotic key rates that can be expected from our experiment. For the key rate calculation we assume error correction at the Shannon limit. We do not provide an error analysis as the resulting key rates are only indicators and not a complete finite size security analysis. 

We show in Fig.~\ref{fig-20keyrate2} how the results of our detailed analysis are comparable to those of Eq.~\ref{eq:keyr}, with the former yielding on average \SI{1722}{bits/s} and the latter approximately \SI{2400}{bits/s}. The key rates could readily be improved by three orders of magnitude through changing the photon source used\cite{PhysRevLett.120.140405}, by improving the detection efficiencies and by optimizing the measurement basis splitting ratios\cite{erven2009entangled}.

To better compare Eq.~\ref{eq:keyr} with our numerical approach, we first take the two-party density matrix computed using the MLM, and remove the vacuum components to form a two-qubit density matrix. We then compute the probabilities for all experimental events using our model POVM elements, and calculate the QBER by Eq.~\ref{eq:QBer}. Then, applying Eq.~\ref{eq:keyr}, we obtain a key rate (squares in Fig.~\ref{fig-20keyrate2}). We see that this method is consistently producing higher key rates (\SI{2110}{bits/s}) than our numerical analysis. We therefore emphasize that both methods use different subsets of the observed data in addition to different proof assumptions. However, the analytical key rate (Eq.~\ref{eq:keyr}) with its assumptions consistently overestimates the key rate in this situation. Nonetheless, the numerical analysis which provides the key rate determined through a physical model permits us to underline the security of the 6 state-4 state scheme.

To further highlight the success of our protocol, we calculate the quality of the entangled two-qubit state shared between Alice and Bob. The calculated mean purity for the experimental data is $0.96(1)$ and $0.94(2)$ for (a) and (b) respectively, while the mean overlap fidelity with a maximally entangled state (Eq.~\ref{eq:bell}) is $0.987(3)$ and $0.980(4)$, and the average concurrence is $0.96(1)$ and $0.94(2)$. An interesting result is that the quality of the measured entangled state is relatively stable throughout the phase variations.

We demonstrate the feasibility of a 6 state-4 state RFI QKD protocol based on entangled photons. Despite the tomographically incomplete set of measurements at one receiver, a secret key can be generated and exchanged between two parties. The security of the protocol is underlined by performing a numerical analysis of the secret key rate. Furthermore, we demonstrate that polarization entangled photons can be reliably transferred, under certain conditions, through high birefringence fibers such as polarization maintaining fibers while preserving the purity of the state.The current pump laser would limit the fiber distance to only a few kilometers. However, this limit can be extended further to distances on the order of hundreds of kilometers by changing to a longer coherence pump laser (coherence times on the order of microseconds). Recently, there have been demonstrations of RFI protocols that reduce the number of states sent to the receiver party\cite{wang2019experimental,liu2019reference}. However, unlike the 6 state-4 state protocol, these protocols do not aim to reduce the required measurement resources. Nonetheless, the 6 state-4 state protocol is also applicable to enhance many other applications, such as polarization compensation systems and quantum LIDAR. Furthermore, the protocol presented here is also applicable for other QKD implementations such as measurement device independent QKD\cite{zhang2019enhanced,wang2017measurement} and for prepare-measure QKD implementations with decoy states, since it can be regarded as a different configuration of the entanglement scheme (i.e. having the entangled source at the sender or receiver), and has the additional configuration flexibility for either the sender or the receiver to perform the 6 state operation. 

\begin{acknowledgments}
	The authors are grateful for the support from the Canada Foundation for Innovation, the Ontario Research Fund, the Canadian Institute for Advanced Research, the Natural Sciences and Engineering Research Council of Canada, the Research Collaboration Funding between Tsinghua University and University of Waterloo, and Industry Canada. The authors thank Dr. Patrick Coles for initial discussions, Kai Hong Li for verifying our calculations, Dr. Sascha Agnes and Karolina S\k{e}dziak-Kacprowicz for assisting building the setup, and Dr. Brendon Higgins and Dr. Jean-Phillipe Bourgeoin for insightful discussions.
\end{acknowledgments}
	
	\nocite{*}
	\raggedright
	\bibliography{tannous64rfi}
	
\end{document}